\documentclass{INTERSPEECH2023}

\interspeechcameraready 

\usepackage{xspace}
\makeatletter
\DeclareRobustCommand\onedot{\futurelet\@let@token\@onedot}
\def\@onedot{\ifx\@let@token.\else.\null\fi\xspace}
\def\eg{\emph{e.g}\onedot} 
\def\ie{\emph{i.e}\onedot}

\makeatother

\usepackage{amsmath,amssymb}

\usepackage{todonotes}

\usepackage{graphicx}
\usepackage{color}
\usepackage{multirow}

\usepackage{colortbl}
\usepackage{array}
\usepackage{nccmath}
\usepackage{booktabs}
\usepackage{tabularx}
\usepackage{algpseudocode}
\usepackage{rotating}
\usepackage{multirow}
\usepackage{comment}

\usepackage[noend]{algorithm2e}
\usepackage{booktabs}
\usepackage{arydshln}
\RestyleAlgo{ruled}

\usepackage[section]{placeins}

\usepackage{url}

\title{A Model for Every User and Budget:\\Label-Free and Personalized Mixed-Precision Quantization}

\name{Edward Fish$^{1,2,*}$\thanks{* Research completed during internship at Samsung Research UK.}, Umberto Michieli$^1$, Mete Ozay$^1$}
\address{
  $^1$Samsung Research UK\\
  $^2$University of Surrey}
\email{edward.fish@surrey.ac.uk, \{u.michieli, m.ozay\}@samsung.com}

\begin{document}

\maketitle
\begin{abstract}
Recent advancement in Automatic Speech Recognition (ASR) has produced large AI models, which become impractical for deployment in mobile devices. Model quantization is effective to produce compressed general-purpose models, however such models may only be deployed to a restricted sub-domain of interest. We show that ASR models can be personalized during quantization while relying on just a small set of unlabelled samples from the target domain. To this end, we propose myQASR, a mixed-precision quantization method that generates tailored quantization schemes for diverse users under any memory requirement with no fine-tuning. myQASR automatically evaluates the quantization sensitivity of network layers by analysing the full-precision activation values. We are then able to generate a personalised mixed-precision quantization scheme for any pre-determined memory budget. Results for large-scale ASR models show how myQASR improves performance for specific genders, languages, and speakers.
The code is available at \url{https://github.com/SamsungLabs/myQASR}.
\end{abstract}

\section{Introduction}

Automatic Speech Recognition (ASR) models improve user experience when interacting with mobile devices while widening the accessibility of such technologies \cite{yu2016automatic,malik2021automatic}. Recent innovation in ASR has focused on large and multi-purpose (\eg, multi-lingual) transformer-based architectures \cite{baevski2020wav2vec,radford2022robust}. However, there has been less consideration for how these models can be effectively compressed and deployed for a diverse range of users and devices, whilst preserving privacy (\ie, with anonymized data).

Model quantization can be divided into three categories \cite{gholami, liang-survey}: i) Quantization Aware Training (QAT) where the quantized model is fine-tuned on labelled data to recover accuracy \cite{nguyen, mishchenko2019low, allenet2022disentangled, ding20224, bie2019simplified, zhen22_interspeech, fasoli22_interspeech}, ii) Post-Training Quantization (PTQ) where labelled data is used to calibrate quantization parameters without training \cite{liu2021post, nagel2020up, hubara2021accurate}, and iii) Data-Free PTQ (DF-PTQ) where data is unlabelled (\ie, label-free) or not available at all \cite{nagel2019data, choi2021qimera}. We focus on label-free PTQ. DF-PTQ has been tackled in the computer vision (CV) domain via layerwise knowledge distillation (KD) between original and quantized layers for compressing CNNs \cite{aghli2021combining, gou2021knowledge, jin2021kdlsq}. These methods have been extended to Transformers for both vision \cite{yuan2022ptq4vit, li2022vit, lin2022fq} and ASR \cite{bie2019simplified,zharikov2022low, wang2022deep, odinokikh2022low}, adjusting for GELU and multi-headed attention. In a deployed scenario, computational and time expensive KD methods are not practical as a copy of the full precision (FP) network is required. When unlabelled data is available, a few methods \cite{liu2021post,li2022q} proposed to minimize distance between both quantized and FP weights and activations, or to minimize a task loss with respect to the quantization parameters. While these methods are promising in CV \cite{li2022q}, their effectiveness has not been explored in ASR. More recently, some data-free methods use statistics of input data to generate synthetic data used to fine-tune quantized models \cite{kim2022integer, yao2022zeroquant}; however, these methods cannot account for out-of-domain data which occur in deployment setups and require further training. In \cite{generative-diverse}, this problem is approached via a diverse sample generation scheme; however, the method still requires fine-tuning on synthetic data, is sensitive to hyper-parameters controlling the skew of sample distributions, and has not been applied to ASR tasks.

Another desirable property of personalized compressed ASR models, is the ability to set any target model size while preserving accuracy. Mixed precision (MP) quantization, where each layer is quantized to a different bit depth, allows the target model size to be controlled. Second-order information \cite{dong2019hawq}, NAS \cite{zhao2021automatic}, and generative methods \cite{cai2020zeroq} are effective for low-bit MP settings in CV, but they either require multiple models to be stored in memory concurrently \cite{zhao2021automatic}, or computationally expensive sensitivity detection \cite{dong2019hawq, cai2020zeroq}. Current MP methods also require several hyper-parameters to search optimum bit-depth combinations or to set the min/max bit depths, which does not permit for fine-grained interpolation between model sizes since extreme values are set \textit{a priori} and layer size is not considered. 

\begin{figure}
    \centering
    \includegraphics[width=0.9\linewidth,trim=0cm 15cm 23.8cm 0cm,clip]{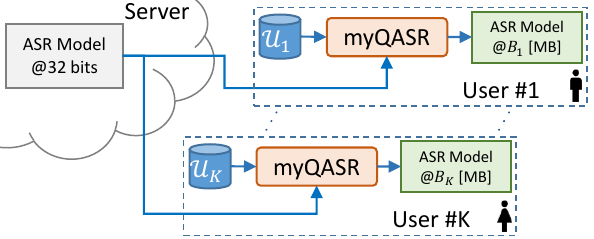}
    \vspace{-0.35cm}
    \caption{Overview of myQASR. A large model is quantized according to users' audio data and their device storage budget.}
    \vspace{-0.6cm}
    \label{fig:GA}
\end{figure}

To address these problems, we present myQASR, a system for personalized compression of ASR models, which performs fast layer-wise sensitivity detection to identify MP bit depths for a range of models (\eg, large multi-purpose transformer models), user traits, and memory constraints.
The scenario targeted by myQASR is depicted in Fig.~\ref{fig:GA}, where a general-purpose FP ASR model is quantized for user $k\in[K]$ given a small dataset of unlabelled private user samples $\mathcal{U}_k$ and target storage budget in MB, $B_k$.
Using $\mathcal{U}_k$, we find a good approximation of weight sensitivity to quantization by observing the median values of FP activations. Our approach is motivated by the activation change among different users (\eg, male and female in Fig.~\ref{fig:motivation}): as such, models for different users require different quantization schemes to find a better compression-accuracy trade-off. 
We then experiment with some calibration methods to find the optimum scale and zero-point for the quantization function given the selected bit depths and statistics of data.

We report experimental validation on recent state-of-the-art architectures (\eg, Wav2Vec2 \cite{baevski2020wav2vec} and Whisper \cite{radford2022robust}) with data segmented according to certain properties (\ie, gender, language, and speaker identity) to demonstrate how our personalized model compression performs over a heterogeneous target.

To summarize, the main contributions of our work are:
1) We introduce myQASR: to our knowledge, the first method for personalized PTQ of ASR models. Our method requires  only a few unlabelled user samples to adjust the quantization parameters without any fine-tuning. %
2) myQASR breaks the common assumption of setting a minimum and a maximum value for the bit depth, and, instead, it relies on a uniformity constraint to guide the quantization process.
3) The uniformity constraint evaluates layer sensitivity in linear time complexity to identify candidate layers that can be further quantized to meet any predefined memory budget constraint to the nearest KB.
4) myQASR is the first ASR quantization approach that quantizes all parts of the network and supports integer bit shifting operations for matrix multiplication to ease on-device deployment.

\vspace{-0.2cm}
\section{Method}
\label{sec:method}
For simplicity, we drop the target user index $k$.
Given a network 
pre-trained 
on a dataset composed of multiple data subsets and parametrized by $\mathcal{W}=\{ W_l \}_{l=1}^{L}$ with $l\in[L]$ layers in the network, we aim to quantize the network to meet any storage budget $B$ minimizing the error rate for a specific target subset $\mathcal{U}$, of which only a few unlabeled samples  are available, with $|\mathcal{U}| \leq 32$ in our experiments.
myQASR can be employed in two stages: 1) layer-wise sensitivity detection - where we perform inference on $\mathcal{U}$ and collect statistics of the raw model, and 2) calibration, where we adjust network parameters based on $\mathcal{U}$.

\begin{figure}[t]
    \centering
    \includegraphics[width=1\linewidth, trim=0.8cm 0.9cm 0.9cm 0.9cm, clip]{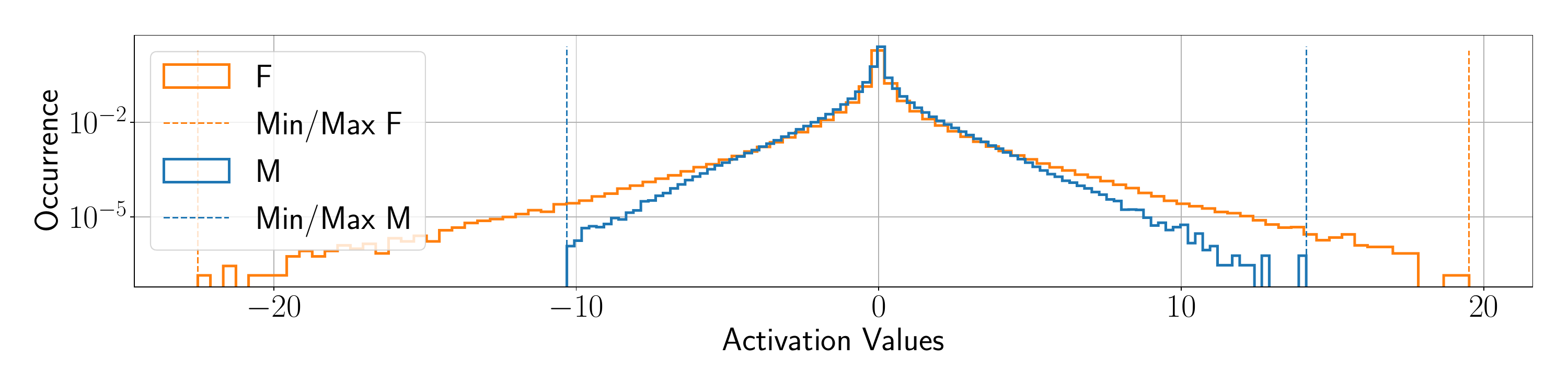}
    \vspace{-0.6cm}
    \caption{Distribution of activations from the first convolution layer of Wav2Vec2 on female (F) and male (M) data.}
    \vspace{-0.2cm}
    \label{fig:motivation}
\end{figure}

\noindent\textbf{Sensitivity Detection.}
To select MP bit-depths $\mathbf{b}\in \mathbb{Z}_+^{L}$, we compute outputs of each Conv and Linear layer\footnote{Self-attention weights (\ie, Queries, Keys, and Values) are considered separately, so they are quantized individually depending on budget.} (\eg, by inserting observers). 
We run inference over the unlabeled target dataset $\mathcal{U}$ storing the median values of output activations, $\mathbf{a} \in \mathbb{R}_+^L$, such that $a_l \triangleq \mathbf{a}[l]$ is the median of outputs obtained at the $l$-th layer $\mathbf{o}_l$. 
We empirically observed a positive correlation between the median of activations and quantization error of the layer with respect to the input samples. As shown in \cite{yu2020low}, at low bit-depths, uniform distribution of weights can reduce quantization error, thus the median measures distribution skew at FP which relates to quantization sensitivity.
Then, we proceed to select the bit depths $\mathbf{b}$ according to the memory budget $B$, as described in Alg.~\ref{alg:mp-select}, where $\lvert W_l \rvert$ counts parameters of $\mathbf{o}_l$.

myQASR performs inference only once with a small number of samples to select the MP scheme, making our method more memory and computationally efficient than distance- or loss- based metrics, such as \cite{dong2019hawq,9772057,liu2021post,yuan2022ptq4vit}. 
Once inference is performed, we take the absolute of median values from each layer (to measure how much they differ from 0 regardless of their sign) and search for the quantization scheme by reducing each layer by one bit until the budget is reached while retaining the highest degree of uniformity among bit depths between layers. 
This uniformity constraint during the sensitivity detection enables removal of any additional hyper-parameters that are required by other methods \cite{chen2021towards, dong2020, li2022vit,cai2020zeroq} which set \textit{a priori} range of bit widths for compression, \ie, $\min(\mathbf{b})$ and $\max(\mathbf{b})$ denoting the minimum and maximum valued bits in $\mathbf{b}$.

\SetKwComment{Comment}{/* }{ */}

\setlength{\textfloatsep}{0.7pt} 

\begin{algorithm}[t]
\caption{Sensitivity detection of myQASR.}
\label{alg:mp-select}
\SetKwFunction{FMain}{ComputeModelSize}
\KwData{$B$ memory budget in MB, $M$ model size in MB ($M>B$), $\mathcal{W}$ model parameters, and $\mathcal{U}$ unlabelled user samples.}
\KwResult{Array $\mathbf{b}$ of selected bit depths.}
$\mathbf{b} \gets \{32, \ldots, 32\}$ 
\ \  \textit{\small// initialize to FP.}\\
Compute median activations $\mathbf{a}$ over $\mathcal{U}$ ($a_l, \forall l \in [L]$)\;
$\hat{\mathbf{q}} \gets$ argsort$(\mathbf{a})$ \ \  \textit{\small// get sorted list of layer indices.}\\

\While{$M > B$}{
  \For{$l$ in $\hat{\mathbf{q}}$ }{
    $b_l -\!= 1$ \ \  \textit{\small// reduce $l$-th layer bit depth by one.} \\
    $M =$ \FMain{$\mathbf{b}$, $\mathcal{W}$} \\
    \lIf{$M <= B$}{
      \Return{bit depth array $\mathbf{b}$} 
    }
    }
}
\vspace{0.1cm}
\DontPrintSemicolon
\SetKwProg{Pn}{def}{:}{\KwRet}
  \Pn{\FMain{$\mathbf{b}$, $\mathcal{W}$}} {
   $\forall (b_l, W_l)$ in $(\mathbf{b}, \mathcal{W})$: \
   qParams += ($b_l$ $/$ $8) \times$ $\lvert W_l \rvert$ \; 
    }
  \KwRet qParams $/$ $1024^2$ \ \ \textit{\small// model size in MB.} \;
  
\end{algorithm}

\noindent\textbf{Quantization \& Calibration.}
Once we have obtained the optimum bit depth, we proceed with quantization. For each layer, we quantize both weights, $W_l$, and inputs, $X_l$, using a quantization function $Q(\theta_l, b_l)$ where $\theta_l \in \{W_l, X_l\}$. The objective is to restrict the FP values of $\theta_l$ to finite integer values by scaling and rounding, as defined by
\vspace{-0.1cm}
\begin{equation}
    Q(\theta_l, b_l) = [\mathrm{round}(\theta_l/S_l) - Z_l]_{b_l} ,
\vspace{-0.1cm}
\label{eq:quantization}
\end{equation} 
where $\mathrm{round}(\cdot)$ is the integer rounding operation, $Z_l$ corrects the zero point of the quantized output, $[\cdot]_{b_l}$ is the representation of $\cdot$ with $b_l$ bits, and $S_l$ is the scaling factor.

In standard uniform quantization \cite{gholami}, $S_l$ is defined by the maximum available values given by $b_l$, \ie, $S_l = 2^{b_l-1}$. 
Uniform quantization function $Q(\theta_l, b_l)$ is directly applicable for quantization of weights $W_l$, as they follow a Gaussian distribution. 
However, 
activations do not \cite{yu2020low} (Fig.~\ref{fig:motivation}), so there may exist $S_l$ values which can minimize the quantization error more effectively. We employ three methods to find appropriate $S_l$ to scale activations, which we describe next.

1) The first method, called myQASR, inserts observers in the network to track the layer-wise minimum ($X^{m}_{l})$ and maximum ($X^{M}_{l})$ values of the input tensors at FP.

The layer scale $S_l$ and zero point $Z_l$ for the input tensor are then obtained by
\vspace{-0.1cm}
 \begin{align}
 S_l = (X^{M}_{l} - X^{m}_{l}) / (2^{b_{l}-1}) , \\ %
 Z_l = -2^{b_{l}-1} - \mathrm{round}(X^{m}_{l} / S_l) .
\vspace{-0.1cm}
\end{align}
We also experiment with other $S_l$, by minimizing the distance between quantized and FP output of layers as described in \cite{easyquant,liu2021post}. In this setting, we define a range of possible values for $S_l$ defined by the minimum and maximum values given by $b_l$, and then take the distance between the FP output (\ie, output vector $\mathbf{o}_l$) and quantized output ($\hat{\mathbf{o}}_l$) where ${\hat{\mathbf{o}}_l = Q(W_l, b_l)^TQ(\mathbf{o}_{l-1}, b_l)}$.
In ablation studies, we evaluate a number of distance metrics for this calibration stage and find that the cosine distance is the most effective.

2) The second method (called myQASR-Hess) is driven by the recent observations \cite{yuan2022ptq4vit} where quantization of GELU and softmax outputs benefits from asymetric non-uniform quantization schemes due to their non-Gaussian distribution. In \cite{yuan2022ptq4vit}, the authors utilise two quantization ranges per layer, $R^l_1$ and $R^l_2$ with scaling factors $S_{R^l_1}$ and $S_{R^l_2}$, where  $R^l_1 = [-2^{k-1}S_{R^l_1}, 0]$ and $R^l_2= [0, -2^{k-1}S_{R^l_2}]$ for post-GELU activations. Finding the optimum $S_{R^l_1}$ and $S_{R^l_2}$ can be performed via a linear search where the objective is to minimize the distance between quantized and FP output of each layer scaled by its impact on the task loss $\mathcal{L}(\hat{y}, y)$. Denoting $\mathbf{\Delta}_l = \hat{\mathbf{o}}_{l} - \mathbf{o}_{l}$, the Hessian-based calibration optimization for layer $l$ is defined by 
\vspace{-0.2cm}
\begin{equation}
\label{eq:hessian}
\displaystyle
    \min_{S_{R^l_1},S_{R^l_2}}\mathbb{E}_{\mathcal{U}}\left[ \mathbf{\Delta}_l^{T} 
    \mathrm{diag} \left[
    \left(\frac{\partial \mathcal{L}_{\mathcal{U}}}{\partial \mathbf{o}_{i}}\right)^{2}
    \right]_{i=1}^{L}
    \mathbf{\Delta}_l\right].
\vspace{-0.2cm}
\end{equation}

3) The third method (myQASR-Cosine) calibrates the model minimizing the cosine distance by
\vspace{-0.2cm}
\begin{equation}
\label{eq:cosine}
\displaystyle
    \min_{S_{R^l_1},S_{R^l_2}}
    \mathbb{E}_{\mathcal{U}}\left[
    \frac{\hat{\mathbf{o}}_l \cdot \mathbf{o}_l}{\lVert \hat{\mathbf{o}}_l \rVert \cdot \lVert \mathbf{o}_l \rVert}
    \right]
    .
\vspace{-0.2cm}
\end{equation}

As we will see, myQASR-Cosine provides the best results for personalized quantization at the cost of a remarked increase in calibration time and memory consumption which may not be available in certain deployment scenarios.

\section{Experimental Analyses}
\noindent\textbf{Datasets.} We employ 3 datasets, one for each personalization task.
To demonstrate our method, we partition data into subsets. However, myQASR requires no \textit{a priori} assumption on data split. We replicate a deployment scenario, where myQASR sees only a small amount of unlabelled target data from a user. 
\\
\textit{1) Gender-wise Personalization.} LibriSpeech (LS) \cite{panayotov2015librispeech} contains $\sim1k$ hours of 16kHz English speech derived from audiobooks.
We perform experiments on \textit{test-clean}, creating Male (M) and Female (F) partitions, splitting audio data by speaker gender.
\textit{2) {Language-wise Personalization}.} FLEURS \cite{fleurs} contains 102 languages, each with $\sim12$ hours of speech. We select $10$ among the top performing languages for the Whisper architecture. 
For each experiment, we randomly sample 32 full spoken sentences for calibration.
\textit{3) Speaker-wise Personalization.}
Google Speech Commands (GSC) \cite{speech-commands} is designed to evaluate effectiveness of keyword spotting models. The dataset contains one second long audio files from multiple speakers. We use the test partition v0.01, which contains 3081 spoken words. For calibration, we select 5 words from each speaker and test performance on all the available test data per speaker.

\noindent\textbf{Models.} 
We use Wav2Vec2 (W2V2) \cite{baevski2020wav2vec} and Whisper \cite{radford2022robust} to evaluate our method. 
For all models, we quantize all weights and activations. On LS, we use a pre-trained W2V2 base (B) model fine-tuned on 960 hours of clean English audio data from LS. 
For Keywords Spotting (KWS), we use a W2V2-Large-Conformer (W2V2-L-C), as described in \cite{fairseq}, pre-trained on GSC. 
Whisper \cite{radford2022robust} is a multi-lingual sequence to sequence (encoder/decoder) Transformer model pre-trained on 680K hours of multi-lingual supervised audio data. We experiment on the Whisper-L, \ie, the large variant. 
We use Word/Character Error Rate (WER/CER) for ASR tasks, and we use accuracy (ACC) for KWS. Average bit depth is denoted by $\bar{b}$.

\begin{figure}[t]
    \centering
    \includegraphics[width=\linewidth,trim=0.4cm 0.5cm 0.4cm 0.3cm,clip]{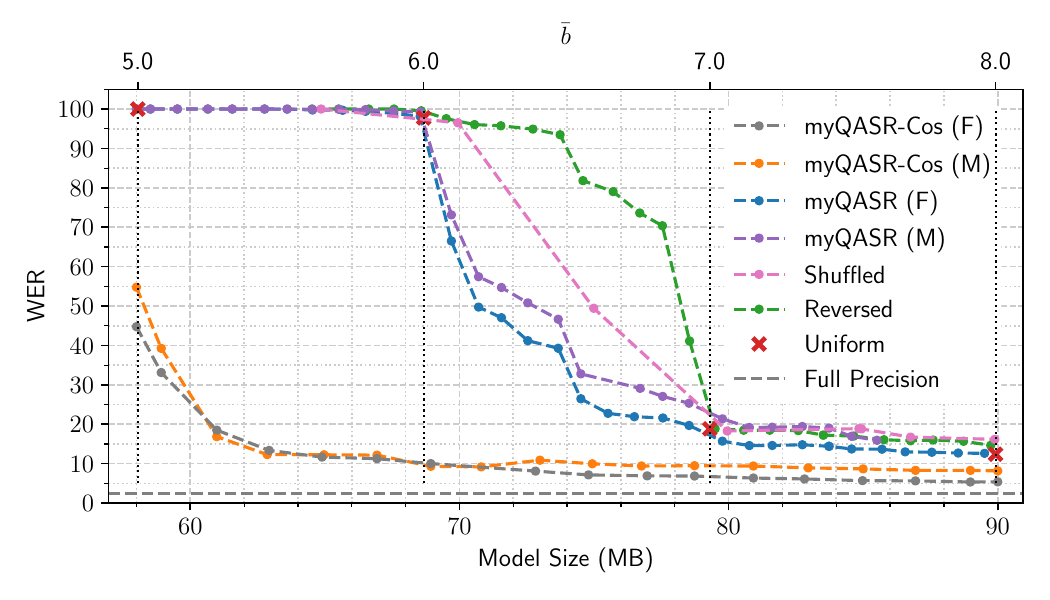}
    \vspace{-0.6cm}
    \caption{WER of W2V2-B on LS-F. Original model is 360MB.}
    \label{fig:w2v2-b-female-main}
    \vspace*{0.3\floatsep}
\newcolumntype{k}[1]{>{\centering\arraybackslash\hspace{0pt}}p{#1}}
    \setlength{\tabcolsep}{0.35pt} %
  \centering
  \vspace{-0cm}
  \resizebox{\linewidth}{!}{%
    \begin{tabular}{k{0.3cm}ck{0.9cm}k{0.9cm}k{0.9cm}k{0.9cm}k{0.9cm}k{0.9cm}k{0.9cm}k{0.9cm}k{0.9cm}k{0.9cm}k{1.2cm}}

        \multicolumn{1}{c}{\multirow{10}[0]{*}{\begin{sideways}Language Test\end{sideways}}} & ca    & \cellcolor[rgb]{ .388,  .745,  .482}8.10 & \cellcolor[rgb]{ .808,  .863,  .506}8.78 & \cellcolor[rgb]{ 1,  .922,  .518}9.12 & \cellcolor[rgb]{ 1,  .922,  .518}9.14 & \cellcolor[rgb]{ .686,  .831,  .498}8.59 & \cellcolor[rgb]{ 1,  .922,  .518}9.10 & \cellcolor[rgb]{ .792,  .859,  .502}8.76 & \cellcolor[rgb]{ 1,  .918,  .518}9.34 & \cellcolor[rgb]{ .78,  .855,  .502}8.74 & \multicolumn{1}{c|}{\cellcolor[rgb]{ 1,  .922,  .518}9.27} & \cellcolor[rgb]{ .973,  .412,  .42}36.00 \\
          & de    & \cellcolor[rgb]{ 1,  .922,  .518}17.38 & \cellcolor[rgb]{ .388,  .745,  .482}17.19 & \cellcolor[rgb]{ .388,  .745,  .482}17.19 & \cellcolor[rgb]{ 1,  .918,  .518}17.65 & \cellcolor[rgb]{ .922,  .898,  .51}17.36 & \cellcolor[rgb]{ .388,  .745,  .482}17.19 & \cellcolor[rgb]{ 1,  .918,  .518}17.74 & \cellcolor[rgb]{ 1,  .918,  .518}17.74 & \cellcolor[rgb]{ 1,  .922,  .518}17.38 & \multicolumn{1}{c|}{\cellcolor[rgb]{ .769,  .855,  .502}17.31} & \cellcolor[rgb]{ .973,  .412,  .42}46.50 \\
          & en    & \cellcolor[rgb]{ 1,  .922,  .518}12.52 & \cellcolor[rgb]{ .949,  .906,  .514}12.45 & \cellcolor[rgb]{ .388,  .745,  .482}11.69 & \cellcolor[rgb]{ 1,  .922,  .518}12.78 & \cellcolor[rgb]{ .949,  .906,  .514}12.45 & \cellcolor[rgb]{ 1,  .922,  .518}12.52 & \cellcolor[rgb]{ 1,  .922,  .518}12.65 & \cellcolor[rgb]{ 1,  .922,  .518}12.52 & \cellcolor[rgb]{ .875,  .882,  .51}12.35 & \multicolumn{1}{c|}{\cellcolor[rgb]{ .827,  .871,  .506}12.29} & \cellcolor[rgb]{ .973,  .412,  .42}75.46 \\
          & fr    & \cellcolor[rgb]{ 1,  .922,  .518}11.85 & \cellcolor[rgb]{ .824,  .871,  .506}11.61 & \cellcolor[rgb]{ 1,  .922,  .518}11.93 & \cellcolor[rgb]{ .388,  .745,  .482}11.02 & \cellcolor[rgb]{ .518,  .78,  .486}11.19 & \cellcolor[rgb]{ 1,  .922,  .518}11.96 & \cellcolor[rgb]{ 1,  .922,  .518}11.93 & \cellcolor[rgb]{ .475,  .769,  .486}11.13 & \cellcolor[rgb]{ .451,  .761,  .482}11.11 & \multicolumn{1}{c|}{\cellcolor[rgb]{ 1,  .91,  .518}12.53} & \cellcolor[rgb]{ .973,  .412,  .42}40.95 \\
          & ja    & \cellcolor[rgb]{ .765,  .851,  .502}14.80 & \cellcolor[rgb]{ .388,  .745,  .482}14.49 & \cellcolor[rgb]{ 1,  .918,  .518}15.15 & \cellcolor[rgb]{ 1,  .922,  .518}15.00 & \cellcolor[rgb]{ .459,  .765,  .486}14.55 & \cellcolor[rgb]{ 1,  .918,  .518}15.18 & \cellcolor[rgb]{ .8,  .863,  .506}14.83 & \cellcolor[rgb]{ 1,  .922,  .518}15.11 & \cellcolor[rgb]{ 1,  .914,  .518}15.30 & \multicolumn{1}{c|}{\cellcolor[rgb]{ .871,  .882,  .51}14.90} & \cellcolor[rgb]{ .973,  .412,  .42}30.56 \\
          & ko    & \cellcolor[rgb]{ .549,  .788,  .49}19.28 & \cellcolor[rgb]{ .729,  .843,  .502}19.46 & \cellcolor[rgb]{ .992,  .761,  .49}21.53 & \cellcolor[rgb]{ 1,  .922,  .518}19.73 & \cellcolor[rgb]{ 1,  .922,  .518}19.73 & \cellcolor[rgb]{ .388,  .745,  .482}19.12 & \cellcolor[rgb]{ .639,  .816,  .494}19.37 & \cellcolor[rgb]{ .996,  .804,  .498}21.08 & \cellcolor[rgb]{ .996,  .827,  .502}20.81 & \multicolumn{1}{c|}{\cellcolor[rgb]{ .906,  .894,  .51}19.64} & \cellcolor[rgb]{ .973,  .412,  .42}25.38 \\
          & nl    & \cellcolor[rgb]{ .875,  .886,  .51}11.70 & \cellcolor[rgb]{ 1,  .922,  .518}11.87 & \cellcolor[rgb]{ .957,  .91,  .514}11.81 & \cellcolor[rgb]{ .549,  .792,  .49}11.23 & \cellcolor[rgb]{ 1,  .91,  .518}12.16 & \cellcolor[rgb]{ 1,  .922,  .518}11.87 & \cellcolor[rgb]{ .388,  .745,  .482}10.99 & \cellcolor[rgb]{ 1,  .898,  .514}12.46 & \cellcolor[rgb]{ .835,  .871,  .506}11.64 & \multicolumn{1}{c|}{\cellcolor[rgb]{ 1,  .922,  .518}11.87} & \cellcolor[rgb]{ .973,  .412,  .42}24.27 \\
          & pl    & \cellcolor[rgb]{ .922,  .898,  .51}12.79 & \cellcolor[rgb]{ .541,  .788,  .49}12.61 & \cellcolor[rgb]{ 1,  .906,  .518}13.47 & \cellcolor[rgb]{ 1,  .91,  .518}13.40 & \cellcolor[rgb]{ .388,  .745,  .482}12.54 & \cellcolor[rgb]{ 1,  .922,  .518}12.93 & \cellcolor[rgb]{ 1,  .922,  .518}12.97 & \cellcolor[rgb]{ .541,  .788,  .49}12.61 & \cellcolor[rgb]{ 1,  .922,  .518}12.82 & \multicolumn{1}{c|}{\cellcolor[rgb]{ .541,  .788,  .49}12.61} & \cellcolor[rgb]{ .973,  .412,  .42}32.67 \\
          & pt    & \cellcolor[rgb]{ 1,  .922,  .518}10.19 & \cellcolor[rgb]{ .631,  .816,  .494}9.91 & \cellcolor[rgb]{ 1,  .922,  .518}9.98 & \cellcolor[rgb]{ .51,  .78,  .486}9.89 & \cellcolor[rgb]{ 1,  .922,  .518}9.98 & \cellcolor[rgb]{ 1,  .922,  .518}10.14 & \cellcolor[rgb]{ 1,  .918,  .518}10.37 & \cellcolor[rgb]{ 1,  .922,  .518}9.98 & \cellcolor[rgb]{ .388,  .745,  .482}9.86 & \multicolumn{1}{c|}{\cellcolor[rgb]{ .875,  .886,  .51}9.96} & \cellcolor[rgb]{ .973,  .412,  .42}37.08 \\
          & ru    & \cellcolor[rgb]{ .816,  .867,  .506}9.62 & \cellcolor[rgb]{ .694,  .831,  .498}9.55 & \cellcolor[rgb]{ .816,  .867,  .506}9.62 & \cellcolor[rgb]{ 1,  .906,  .518}10.09 & \cellcolor[rgb]{ .447,  .761,  .482}9.42 & \cellcolor[rgb]{ .388,  .745,  .482}9.38 & \cellcolor[rgb]{ 1,  .914,  .518}9.96 & \cellcolor[rgb]{ 1,  .906,  .518}10.12 & \cellcolor[rgb]{ 1,  .886,  .514}10.49 & \multicolumn{1}{c|}{\cellcolor[rgb]{ 1,  .922,  .518}9.72} & \cellcolor[rgb]{ .973,  .412,  .42}20.28 \\
          &       & ca    & de    & en    & fr    & ja    & ko    & nl    & pl    & pt    & \multicolumn{1}{c|}{ru} & No Calib \\
          &       & \multicolumn{11}{c}{Language Calibration} \\

    \end{tabular}%
    }
    \vspace{-0.4cm}
  \caption{WER on FLEURS with myQASR-Whisper-L.}
  \label{fig:whisper-lang}%
\vspace*{0.2\floatsep}

    \setlength{\tabcolsep}{1.6pt} %
  \centering
  \vspace{-0cm}
  \resizebox{\linewidth}{!}{%
    \begin{tabular}{k{0.05cm}ck{0.8cm}k{0.8cm}k{0.8cm}k{0.8cm}k{0.8cm}k{0.8cm}k{0.8cm}k{0.8cm}k{0.8cm}k{0.8cm}k{1.2cm}}
    
    \multicolumn{1}{c}{\multirow{10}[0]{*}{\begin{sideways}Speaker ID Test\end{sideways}}} & \multicolumn{1}{c}{1} & \cellcolor[rgb]{ .388,  .745,  .482}90.9 & \cellcolor[rgb]{ .388,  .745,  .482}90.9 & \cellcolor[rgb]{ .388,  .745,  .482}90.9 & \cellcolor[rgb]{ .973,  .412,  .42}81.8 & \cellcolor[rgb]{ .388,  .745,  .482}90.9 & \cellcolor[rgb]{ .973,  .412,  .42}81.8 & \cellcolor[rgb]{ .388,  .745,  .482}90.9 & \cellcolor[rgb]{ .388,  .745,  .482}90.9 & \cellcolor[rgb]{ .388,  .745,  .482}90.9 & \multicolumn{1}{k{0.8cm}|}{\cellcolor[rgb]{ .973,  .412,  .42}81.8} & \cellcolor[rgb]{ .388,  .745,  .482}90.9 \\
          & \multicolumn{1}{c}{2} & \cellcolor[rgb]{ .984,  .663,  .467}78.6 & \cellcolor[rgb]{ .388,  .745,  .482}100 & \cellcolor[rgb]{ .388,  .745,  .482}100 & \cellcolor[rgb]{ 1,  .922,  .518}85.7 & \cellcolor[rgb]{ .984,  .663,  .467}78.6 & \cellcolor[rgb]{ 1,  .922,  .518}85.7 & \cellcolor[rgb]{ .388,  .745,  .482}100 & \cellcolor[rgb]{ .984,  .663,  .467}78.6 & \cellcolor[rgb]{ .984,  .663,  .467}78.6 & \multicolumn{1}{c|}{\cellcolor[rgb]{ .973,  .412,  .42}71.4} & \cellcolor[rgb]{ .694,  .835,  .502}92.9 \\
          & \multicolumn{1}{c}{3} & \cellcolor[rgb]{ 1,  .922,  .518}91.7 & \cellcolor[rgb]{ 1,  .922,  .518}91.7 & \cellcolor[rgb]{ .388,  .745,  .482}100 & \cellcolor[rgb]{ 1,  .922,  .518}91.7 & \cellcolor[rgb]{ 1,  .922,  .518}91.7 & \cellcolor[rgb]{ 1,  .922,  .518}91.7 & \cellcolor[rgb]{ .973,  .412,  .42}83.3 & \cellcolor[rgb]{ 1,  .922,  .518}91.7 & \cellcolor[rgb]{ 1,  .922,  .518}91.7 & \multicolumn{1}{c|}{\cellcolor[rgb]{ .973,  .412,  .42}83.3} & \cellcolor[rgb]{ 1,  .922,  .518}91.7 \\
          & \multicolumn{1}{c}{4} & \cellcolor[rgb]{ 1,  .922,  .518}52.6 & \cellcolor[rgb]{ .957,  .91,  .518}54.4 & \cellcolor[rgb]{ .675,  .827,  .502}64.9 & \cellcolor[rgb]{ .388,  .745,  .482}75.4 & \cellcolor[rgb]{ .973,  .412,  .42}45.6 & \cellcolor[rgb]{ .992,  .792,  .49}50.9 & \cellcolor[rgb]{ 1,  .922,  .518}52.6 & \cellcolor[rgb]{ .984,  .667,  .467}49.1 & \cellcolor[rgb]{ .992,  .792,  .49}50.9 & \multicolumn{1}{c|}{\cellcolor[rgb]{ .973,  .412,  .42}45.6} & \cellcolor[rgb]{ .863,  .882,  .51}57.9 \\
          & \multicolumn{1}{c}{5} & \cellcolor[rgb]{ 1,  .922,  .518}83.3 & \cellcolor[rgb]{ 1,  .922,  .518}83.3 & \cellcolor[rgb]{ .388,  .745,  .482}91.7 & \cellcolor[rgb]{ 1,  .922,  .518}83.3 & \cellcolor[rgb]{ .388,  .745,  .482}91.7 & \cellcolor[rgb]{ .973,  .412,  .42}75.0 & \cellcolor[rgb]{ .388,  .745,  .482}91.7 & \cellcolor[rgb]{ 1,  .922,  .518}83.3 & \cellcolor[rgb]{ .388,  .745,  .482}91.7 & \multicolumn{1}{c|}{\cellcolor[rgb]{ 1,  .922,  .518}83.3} & \cellcolor[rgb]{ .996,  .918,  .514}83.3 \\
          & \multicolumn{1}{c}{6} & \cellcolor[rgb]{ 1,  .922,  .518}93.3 & \cellcolor[rgb]{ .388,  .745,  .482}100 & \cellcolor[rgb]{ .388,  .745,  .482}100 & \cellcolor[rgb]{ 1,  .922,  .518}93.3 & \cellcolor[rgb]{ .984,  .667,  .467}86.7 & \cellcolor[rgb]{ .388,  .745,  .482}100 & \cellcolor[rgb]{ .388,  .745,  .482}100 & \cellcolor[rgb]{ .984,  .667,  .467}86.7 & \cellcolor[rgb]{ 1,  .922,  .518}93.3 & \multicolumn{1}{c|}{\cellcolor[rgb]{ .973,  .412,  .42}80.0} & \cellcolor[rgb]{ .996,  .918,  .514}93.3 \\
          & \multicolumn{1}{c}{7} & \cellcolor[rgb]{ 1,  .922,  .518}75.0 & \cellcolor[rgb]{ 1,  .922,  .518}75.0 & \cellcolor[rgb]{ .596,  .808,  .498}87.5 & \cellcolor[rgb]{ .984,  .667,  .467}68.8 & \cellcolor[rgb]{ 1,  .922,  .518}75.0 & \cellcolor[rgb]{ .973,  .412,  .42}62.5 & \cellcolor[rgb]{ .388,  .745,  .482}93.8 & \cellcolor[rgb]{ .984,  .667,  .467}68.8 & \cellcolor[rgb]{ .973,  .412,  .42}62.5 & \multicolumn{1}{c|}{\cellcolor[rgb]{ 1,  .922,  .518}75.0} & \cellcolor[rgb]{ .8,  .867,  .51}81.3 \\
          & \multicolumn{1}{c}{8} & \cellcolor[rgb]{ .984,  .667,  .467}50.0 & \cellcolor[rgb]{ .694,  .835,  .502}80.0 & \cellcolor[rgb]{ .694,  .835,  .502}80.0 & \cellcolor[rgb]{ 1,  .922,  .518}60.0 & \cellcolor[rgb]{ .973,  .412,  .42}40.0 & \cellcolor[rgb]{ 1,  .922,  .518}60.0 & \cellcolor[rgb]{ .694,  .835,  .502}80.0 & \cellcolor[rgb]{ .388,  .745,  .482}100 & \cellcolor[rgb]{ 1,  .922,  .518}60.0 & \multicolumn{1}{c|}{\cellcolor[rgb]{ 1,  .922,  .518}60.0} & \cellcolor[rgb]{ 1,  .922,  .518}60.0 \\
          & \multicolumn{1}{c}{9} & \cellcolor[rgb]{ 1,  .922,  .518}73.3 & \cellcolor[rgb]{ .984,  .667,  .467}66.7 & \cellcolor[rgb]{ .388,  .745,  .482}80.0 & \cellcolor[rgb]{ 1,  .922,  .518}73.3 & \cellcolor[rgb]{ 1,  .922,  .518}73.3 & \cellcolor[rgb]{ 1,  .922,  .518}73.3 & \cellcolor[rgb]{ .973,  .412,  .42}60.0 & \cellcolor[rgb]{ 1,  .922,  .518}73.3 & \cellcolor[rgb]{ .388,  .745,  .482}80.0 & \multicolumn{1}{c|}{\cellcolor[rgb]{ 1,  .922,  .518}73.3} & \cellcolor[rgb]{ .973,  .412,  .42}60.0 \\
          & \multicolumn{1}{c}{10} & \cellcolor[rgb]{ 1,  .922,  .518}75.0 & \cellcolor[rgb]{ .984,  .667,  .467}66.7 & \cellcolor[rgb]{ .388,  .745,  .482}91.7 & \cellcolor[rgb]{ 1,  .922,  .518}75.0 & \cellcolor[rgb]{ .973,  .412,  .42}58.3 & \cellcolor[rgb]{ 1,  .922,  .518}75.0 & \cellcolor[rgb]{ 1,  .922,  .518}75.0 & \cellcolor[rgb]{ 1,  .922,  .518}75.0 & \cellcolor[rgb]{ .984,  .667,  .467}66.7 & \multicolumn{1}{c|}{\cellcolor[rgb]{ .388,  .745,  .482}91.7} & \cellcolor[rgb]{ 1,  .922,  .518}75.0 \\
          &       & 1     & 2     & 3     & 4     & 5     & 6     & 7     & 8     & 9     & \multicolumn{1}{c|}{10} & No Calib \\
          &       & \multicolumn{11}{c}{Speaker ID Calibration} \\
    \end{tabular}%
    }
    \vspace{-0.4cm}
    \caption{KWS ACC on GSC with myQASR-W2V2-L-C.} 
   \label{fig:w2v2-l-id-conf}%
\end{figure}

\begin{table*}[ht!]
    \centering
   \setlength{\tabcolsep}{1.1pt}
\footnotesize
  \centering
  \caption{Ablation on min-max (mm) MP bit depths selection. Size: in MB and min-max values within brackets (min-max), my: myQASR.}
  \vspace{-0.3cm}
  \resizebox{\linewidth}{!}{%
    \begin{tabular}{lcccccccccccccccccc}
    \toprule
          & size & WER$_{\mathrm{mm}}$ & WER$_\mathrm{my}$ & size & WER$_{\mathrm{mm}}$ & WER$_\mathrm{my}$ & size & WER$_{\mathrm{mm}}$ & WER$_\mathrm{my}$ & size & WER$_{\mathrm{mm}}$ & WER$_\mathrm{my}$ & size & WER$_{\mathrm{mm}}$ & WER$_\mathrm{my}$ & size & WER$_{\mathrm{mm}}$ & WER$_\mathrm{my}$ \\\cmidrule(lr){2-4}\cmidrule(lr){5-7}\cmidrule(lr){8-10}\cmidrule(lr){11-13}\cmidrule(lr){14-16}\cmidrule(lr){17-19}
    \textbf{M} &  82.5 (5-7)     & 6.6      &  4.7     &    87.7 (6-7)   &   5.6    &   4.3    &  81.9 (4-8)      & 6.9      &  6.6     &  87.5 (5-8)     &   6.4    &   4.3    &  93.2 (6-8)     &  5.4     &  4.1     &    98.4 (7-8)   &   4.2    &  4.2\\
    \textbf{F} & 82.3 (5-7) & 7.4  & 5.3  & 97.7 (6-7) & 5.5  & 4.9  &  82.1(4-8) & 7.1 & 6.2  & 87.5 (5-8) & 7.0  & 4.9  & 93.0 (6-8) & 5.3  & 4.7  & 98.2 (7-8) & 4.6  & 4.6 \\
    \bottomrule
    \end{tabular}%
    }
    \vspace{-0.55cm}
  \label{tab:ablation_mp}%
\end{table*}

\noindent\textbf{Main Results on Gender.} Fig.~\ref{fig:w2v2-b-female-main} shows WER of W2V2-B pre-trained on multi-gender data, quantized for a female user and tested on LS-F. %
The plot spans an increasing memory budget from 60MB (\ie, $\bar{b}\approx5$) to 90MB (\ie, $\bar{b}\approx8$).
Original (\ie, FP, non-quantized) model performance indicates the lower bound for WER (gray dashed line).
Uniform quantization yields competitive results, however cannot meet fine-grained memory requirements.
Our simplest myQASR calibrated on female data, myQASR (F), improves accuracy and can meet any desired target model size. 
We argue on the usefulness of our sensitivity method since: 
myQASR (F) shows significant benefits compared to myQASR (M), \ie, quantizing the model according to male data; 
myQASR (F) outperforms  its shuffled (\ie, bit-depths shuffled) or reversed (\ie, bit-depths reversed) versions by large margin. 
Cosine-based calibration brings large benefits, reducing the gap from the FP model. Nonetheless, calibration on female data still outperforms calibration on male data.

\noindent\textbf{Main Results on Language.}
In Fig.~\ref{fig:whisper-lang}, we show personalized compression in multi-lingual settings. We take the pre-trained multi-lingual Whisper-L model and calibrate bit-depths and activation ranges using just 32 samples of unlabelled data. Each language label represents a tune and test split, and we show that calibrating bit depths and activations for the same language leads to improved results.
Although we obtain better results on the same language used for calibration (on-diagonal results), we remark that the resulting model still achieves competitive results on other languages (off-diagonal results); thus being able to predict also on such languages. 
In the worst case (\ie, Russian), our method is outperformed by calibration on other languages. However, it shows a relative gain of $0.9\%$ compared to the average of other-language results.
In the best case (\ie, Catalan), our method outperforms the average of other-language by $10.9\%$ relative gain.
On average, our same-language myQASR yields 66.2\% better results than standard uniform quantization with no calibration (12.5\% vs.\ 36.9\% WER), and 4.2\% better results than other-language quantization (13.0\% WER).

\noindent\textbf{Main Results on Speaker.}
In Fig.~\ref{fig:w2v2-l-id-conf}, we show ACC for our myQASR applied to a W2V2-L-C \cite{fairseq} compressed from 2.4GB to 375MB (\ie, $\bar{b}=5\mathrm{bits}$). We partition GSC by speaker ID and evaluate on each ID with calibration data from different speakers. We show that, when sensitivity and calibration analysis is performed on the same speaker, we achieve optimum performance. For example, in the best case (\ie, speaker $\#7$), we achieve $100\%$ ACC when compression is personalized for that speaker, compared with $40\%$ ACC when personalized for another speaker from the same dataset, even though keywords are the same.
On average, our same-speaker myQASR yields 17.5\% higher results than standard uniform quantization with no calibration (92.3\% vs. 78.6\% ACC), and 19.6\% higher results than other-speaker quantization (77.2\% ACC).

\section{Ablation Study}
Ablation is performed on W2V2-B compressed via myQASR-Cosine to 75MB, i.e $\bar{b}=6.5$, (unless otherwise stated) on gender data calibrated and tested on the same split. 

\noindent\textbf{Bit Depth Selection \& Uniformity Constraint.}
Tab.~\ref{tab:ablation_mp} shows a comparison between our uniform constraint and the common min-max method \cite{dong2019hawq, cai2020zeroq}, where min-max values are chosen for depths according to a linear interpolation mapping highest (lowest) activation value to the min (max) bit depth. This approach leads to significantly lower results at fixed target compression ratios than ours. This shows the advantage of enforcing some uniformity among layers in the network. Using min-max bit depths also requires two more hyper-parameters that we avoid thanks to the sensitivity evaluation scheme, as discussed next.

\noindent\textbf{Sensitivity} 
is evaluated in Tab.~\ref{tab:ablation_sensitivity}.
We grouped methods as reduction-based or distance-based.
Reduction-based methods compute an aggregate measure of the distribution of activations obtained using the original model, namely: average, median (ours), max, max of the absolute and standard deviation.
Distance-based methods compute a distance measure between layer-wise activations obtained using the quantized and original model, namely: L1, L2, Spectral norm, Frobenius norm, and KL divergence. As in the reduction setting, the values are sorted in order of increasing distance and then used to assign bit depths per layer. 
Reduction-based methods are more practical than distance-based ones as they do not require both the quantized and original model, and can achieve performance comparable to distance-based methods.
Among reduction-based approaches, median provides the best results. We reason that the median provides a measure of distribution skew at FP which correlates with quantization sensitivity as described in \cite{yu2020low}.

\noindent\textbf{Calibration} is evaluated in Tab.~\ref{tab:ablation_calibration}. To evaluate it, we use a number of distance functions which compute the quantization error between FP and quantized activations. For each metric, apart from myQASR variations, we generate $t$ possible quantization scales per each layer and minimize the error defined by the distance metric ($t=100$, as in \cite{yuan2022ptq4vit}). Cosine and Hessian weighted myQASR perform the best but have a high computational search time. 
L1 achieves competitive results, however its WER is outperformed by myQASR-Hess with a similar calibration time. 
myQASR, which performs a simple min/max calibration as described in Sec.~\ref{sec:method}, provides a trade-off between accuracy and computational time, and does not require linear search or any additional hyper-parameters as the other methods. 

We study the amount of unlabelled target data needed in Tab.~\ref{tab:ablation_num_samples} and verify that a few samples are sufficient for calibration, as reported in \cite{nni_link_observer_num_samples}. For robustness, we choose $32$ samples, since variability of results is minimized (\ie, low standard deviation). Similarly, we perform personalized compression via $5$ samples for speakers and $32$ for languages.

\begin{table}[t]
  \centering
  \footnotesize
  \setlength{\tabcolsep}{1.5pt}
  \caption{Ablation on the sensitivity scheme. We take only the measures used in compared methods for a fair comparison.}
  \vspace{-0.3cm}
  \resizebox{\linewidth}{!}{%
    \begin{tabular}{lcccc:lcccc}
    \toprule
    \multicolumn{1}{c}{\multirow{2}[0]{*}{\rotatebox{1.35}{\textbf{Reduction}}}} & \multicolumn{2}{c}{\textbf{Male}} & \multicolumn{2}{c}{\textbf{Female}} & \multicolumn{1}{:c}{\multirow{2}[0]{*}{\rotatebox{1.35}{\textbf{Distance}}}} & \multicolumn{2}{c}{\textbf{Male}} & \multicolumn{2}{c}{\textbf{Female}} \\
    \cmidrule(lr){2-3}\cmidrule(lr){4-5}\cmidrule(lr){7-8}\cmidrule(lr){9-10}
          & \textbf{WER} & \textbf{CER} & \textbf{WER} & \textbf{CER} &       & \textbf{WER} & \textbf{CER} & \textbf{WER} & \textbf{CER} \\
          \midrule
    Avg  & 7.5  & 2.4  &  7.3 & 2.3  & L1 \cite{yuan2022ptq4vit} & 7.3  & 2.3  & 7.2  & \textbf{2.2} \\
    Median (ours) & \textbf{6.6} & \textbf{2.1} & \textbf{7.1} & \textbf{2.2} & L2 \cite{yuan2022ptq4vit} & 7.2  & 2.2  & 7.2  & \textbf{2.2} \\
    Max  & 7.1  & 2.2  & 7.8  & 2.4  & SN \cite{liu2021post} & 7.3  & 2.3  & 7.3  & \textbf{2.2} \\
    Max Abs  & 7.2  & 2.3  & 8.4  & 2.7  & Frob \cite{9772057} & 7.3 & 2.3  & 7.2  & \textbf{2.2} \\
    Std  & 7.5  & 2.4  & 7.4  & 2.3  & KL \cite{yuan2022ptq4vit} & 7.3  & 2.3  & 8.4  & 2.7 \\
    \bottomrule
    \end{tabular}%
    }
  \label{tab:ablation_sensitivity}%
  \vspace*{0.5\floatsep}
\setlength{\tabcolsep}{4.5pt}
\footnotesize
  \centering
  \caption{Ablation on the calibration scheme. Time (sec) measures calibration procedure only.}
  \vspace{-0.3cm}
    \resizebox{0.9\linewidth}{!}{%
    \begin{tabular}{lcccccc}
    \toprule
    \multirow{2}[0]{*}{} & \multicolumn{3}{c}{\textbf{Male}} & \multicolumn{3}{c}{\textbf{Female}} \\\cmidrule(lr){2-4}\cmidrule(lr){5-7}
          & \multicolumn{1}{c}{\textbf{WER}} & \multicolumn{1}{c}{\textbf{CER}} & \multicolumn{1}{c}{\textbf{Time}} & \multicolumn{1}{c}{\textbf{WER}} & \multicolumn{1}{c}{\textbf{CER}} & \multicolumn{1}{c}{\textbf{Time}} \\
          \midrule
    None & 87.5 & 67.2 & 0 &  66.2 & 83.0 & 0 \\
    L1 \cite{yuan2022ptq4vit} & 8.2  & 2.7  & 158   & 9.8  & 2.8  & 147 \\
    L2 \cite{yuan2022ptq4vit} & 63.7 & 3.4 & 155   & 57.5 & 29.4 & 156 \\
    LinW L2 \cite{yuan2022ptq4vit} & 89.5 & 62.9 & 154   & 92.2 & 70.2 & 155 \\
    SqW L2 \cite{yuan2022ptq4vit} & 94.7 & 81.0 & 154   & 97.1 & 82.9 & 155 \\
    myQASR & 28.8 & 28.7 & \textbf{7} & 22.7 & 22.6 & \textbf{7} \\
    myQASR-Hess & 7.9  & 2.5  & 161   & 7.9  & 2.4  & 154 \\
    myQASR-Cosine & \textbf{6.6} & \textbf{2.1} & 172 & \textbf{7.1} & \textbf{2.2} & 171 \\
    \bottomrule
    \end{tabular}%
    }
  \label{tab:ablation_calibration}%
  
    \vspace*{0.5\floatsep}
\setlength{\tabcolsep}{2pt}
\footnotesize
  \centering
  \caption{Ablation on $|\mathcal{U}|$, results averaged over 5 seeds.}
  \vspace{-0.3cm}
  \resizebox{\linewidth}{!}{%
    \begin{tabular}{lcccccc}
    \toprule
    \textbf{$|\mathcal{U}|$} & \textbf{4} & \textbf{8} & \textbf{16} & \textbf{32} & \textbf{64} & \textbf{128} \\ %
    \midrule
    \textbf{WER (M)} &  \textbf{6.1}$\pm$2.0    & 7.0$\pm$0.4      &  6.9$\pm$0.8  & 6.6$\pm$\textbf{0.1}   & 8.4$\pm$0.8     & 8.5$\pm$1.4  \\
    \textbf{WER (F)} & 7.1$\pm$1.6  & 7.0$\pm$1.0  & \textbf{6.6}$\pm$1.0  & 7.1$\pm$\textbf{0.2} & 6.9$\pm$0.8  & 6.8$\pm$0.5
    \\\bottomrule
    \end{tabular}%
    }
  \label{tab:ablation_num_samples}%
\vspace*{0.5\floatsep}
\setlength{\tabcolsep}{1.5pt}
\footnotesize
  \centering
  \caption{Ablation on activation quantization.}
  \vspace{-0.3cm}
    \begin{tabular}{lcccccccccccccc}
    \toprule
          & \multicolumn{7}{c}{\textbf{Male}}              & \multicolumn{7}{c}{\textbf{Female}} \\\cmidrule(lr){2-8}\cmidrule(lr){9-15}
    \textbf{Act bits} & \textbf{4} & \textbf{6} & \textbf{8} & \textbf{10} & \textbf{12} & \textbf{16} & \textbf{32} &\textbf{4} & \textbf{6} & \textbf{8} & \textbf{10} & \textbf{12} & \textbf{16} & \textbf{32} \\
    \textbf{WER} & 69.0 & 8.1 &  \textbf{6.6}  & 7.0  & 7.0 & 7.1& 7.1& 70.0 & 8.5  & \textbf{7.1}  & 7.5 & 7.4 & 7.4 & 7.4\\
    \textbf{CER} & 34.7 & 2.7  & 2.2  & 2.2  & 2.2 & \textbf{2.1} & \textbf{2.1} & 3.6 & 2.7  & \textbf{1.9}  & 2.3  & 2.3 & 2.3 & 2.3 \\
    \bottomrule
    \end{tabular}%
  \label{tab:ablation_activation}%
  \vspace{0.2cm}
\end{table}%

\noindent\textbf{Activation} quantization is evaluated in Tab.~\ref{tab:ablation_activation} (also weights are quantized). As expected, we observe performance improvements when quantization decreases. We select 8-bit quantization following previous approaches \cite{9772057} and verify that it represents a good trade-off between compression and accuracy. We note how female-specific models are more sensitive to quantization 
which is a reflection of an original bias of the FP model.

\section{Conclusions}

We introduced a new task of personalized model quantization to bring large ASR transformers on low-resource devices such as mobile and edge devices with performance targeted for the final end user. To address this, we propose myQASR, a versatile personalized quantization scheme to compress large ASR models to any memory budget. myQASR couples a uniformity constraint to evaluate layer sensitivity with optional Hessian guidance to set quantization scaling parameters. It requires only a few user-specific unlabelled samples to drive the quantization process personalizing the model performance with no fine-tuning.
Our work provides a baseline for future research in personalized compression focusing on greater accessibility and performance for diverse users and devices.

\bibliographystyle{IEEEtran}
\bibliography{mybib}

\begin{center}
  \textbf{\large A Model for Every User and Budget:\\Label-Free and Personalized Mixed-Precision Quantization \\ \vspace{0.1cm} \textit{Supplementary Material}}
\end{center} 

\setcounter{equation}{0}
\setcounter{figure}{0}
\setcounter{table}{0}
\setcounter{page}{1}
\setcounter{section}{0}
\renewcommand{\theequation}{S\arabic{equation}}
\renewcommand{\thefigure}{S\arabic{figure}}
\renewcommand{\thetable}{S\arabic{table}}
\renewcommand{\thesection}{S\arabic{section}}

\noindent In this document we report some additional results to support our main findings.
In Sec.~\ref{sec:data_stats} we present the statistics of the three personalization datasets employed in our work: namely, gender- language- and speaker-personalization.
In Sec.~\ref{sec:details} we include additional details on implementations and hyper-parameters.
In Sec.~\ref{sec:gender} we discuss additional results of Wav2Vec2-Base (W2V2-B) on male data and of Wav2Vec2-Large (W2V2-L) on both genders.
Sec.~\ref{sec:w2v2l} discusses ablation results on W2V2-L.
Finally, Sec.~\ref{sec:qualitative} reports some model transcriptions.

\section{Datasets Statistics}
\label{sec:data_stats}

\noindent\textbf{Gender-Personalization via LibriSpeech.}
LibriSpeech (LS) \cite{panayotov2015librispeech} contains $\sim1k$ hours of 16kHz English speech derived from audiobooks from the LibriVox project, and has been carefully segmented and aligned. We perform experiments on \textit{test-clean}, and create Male (M) and Female (F) partitions, splitting audio data by speaker gender, using the labels provided. We report the number of samples in the testing set in Tab.~\ref{tab:librispeech_distribution}. We notice how each gender is a representation of multiple speakers, therefore, myQASR is not personalized to a specific speaker, but rather to the gender itself.

\begin{figure}[t]
    \centering
    \includegraphics[width=\linewidth]{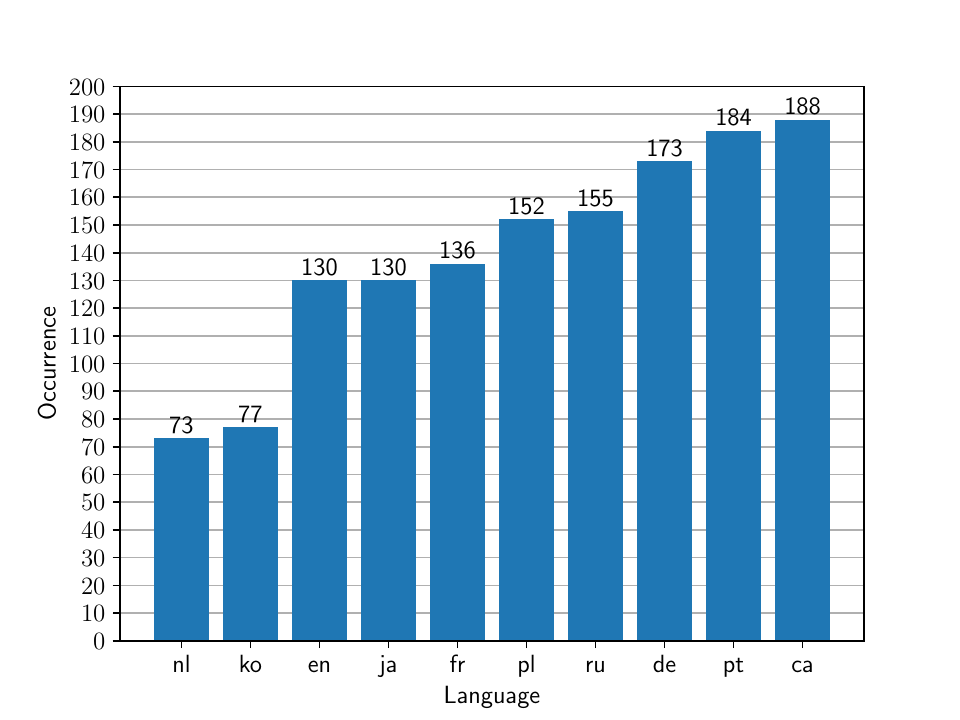}
    \caption{Number of test samples in FLEURS for each language.}
    \label{fig:fleurs_distribution}
\end{figure}

\noindent\textbf{Language-Personalization via FLEURS.} FLEURS \cite{fleurs} contains 102 languages, each with $\sim 12$ hours of speech. We select $10$ among the top performing languages for the Whisper architecture (\ie, Korean \textit{ko}, English \textit{en}, Japanese \textit{ja}, Dutch \textit{nl}, Portuguese \textit{pt}, German \textit{de}, Polish \textit{pl}, Russian \textit{ru}, Catalan \textit{ca}, and French \textit{fr}). Fig.~\ref{fig:fleurs_distribution} reports the distribution of samples in the testing set. Our method can personalize the quantized model to find the most effective bit and calibration setup for each language using only a few samples for sensitivity detection and calibration (\ie, in the results we use 32 samples).

\begin{figure}[!ht]
    \centering
    \includegraphics[width=\linewidth]{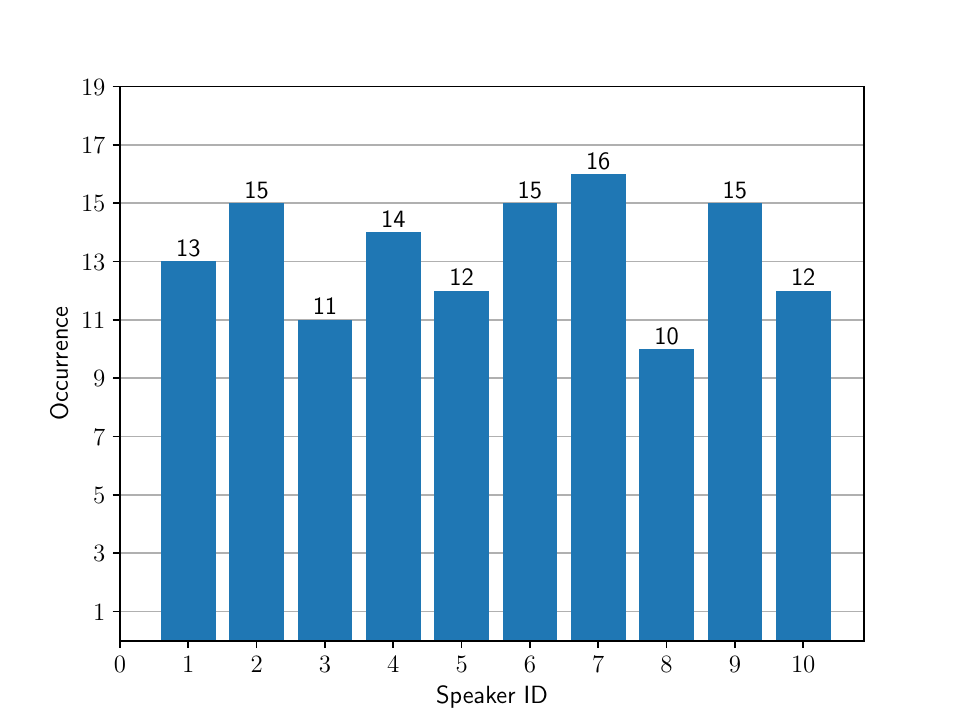}
    \caption{Number of test samples per speaker of Google Speech Commands. Speaker IDs are shown as reported in main paper. Corresponding ID's are shown in Sec.~\ref{sec:data_stats}}
    \label{fig:gsc}
\end{figure}

\noindent\textbf{Speaker-Personalization via GSC.}
Google Speech Commands (GSC) \cite{speech-commands} dataset is designed for the evaluation of keyword spotting models. The dataset contains one second long audio files from multiple speakers. We use the test partition v0.01 which contains 3081 spoken words. For calibration, we select 5 words from each speaker and test performance on all the available test data per speaker. We report the distribution of samples per speaker in Fig.~\ref{fig:gsc} and evaluate on 10 speakers. We select all speakers with more than 10 utterances and then take a random sample of 10. The ID's in the paper correspond to the following speaker ID's. $1:\textit{6205088b}$, $2:\textit{d5ca80c6}$, $3:\textit{3ff840aa}$,  $4:\textit{a60a09cf}$, $5:\textit{f292725f
}$, $6:\textit{105a0eea}$, $7:\textit{d7467392}$, $8:\textit{95ba4996}$, $9:\textit{26b28ea7}$, $10:\textit{90b94017}$.

\begin{table}[t]
\setlength{\tabcolsep}{9pt}
\footnotesize
  \centering
  \caption{Number of speakers and samples in the test partition of LibriSpeech.}
  \vspace{0cm}
    \begin{tabular}{lcc}
    \toprule
    \textbf{Gender} & \textbf{Speakers} & \textbf{Samples} \\
    \midrule
    \textbf{Male} & 38 & 1389 \\
    \textbf{Female} & 49 & 1231\\
    \midrule
    \textbf{Total} & 87 & 2620\\
   
    \bottomrule
    \end{tabular}%
  \label{tab:librispeech_distribution}%
  \vspace{0cm}
\end{table}

\vspace{3cm}
\section{Additional Details}
\label{sec:details}

\subsection{Implementation Details}
All baselines are implemented in PyTorch on a single codebase to guarantee fair comparisons. The results are evaluated on a NVIDIA GeForce RTX 3090 GPU supported by an Intel(R) Xeon(R) Gold 5218R CPU @ 2.10GHz with 80 cores and 252GB RAM.

\subsection{Hyper-parameters}

For myQASR-Hess and myQASR-Cosine we set all hyper-parameters as described in \cite{yuan2022ptq4vit}. The number of candidate values for $S_l$ is set to 100 and we perform 3 search rounds where we alternate between $S_{R_1^l}$ and $S_{R_2^l}$. For calibration, we perform parallel quantization as described in \cite{yuan2022ptq4vit} in which we search for optimum $S_{R_1^l}$ and $S_{R_2^l}$ while $\textbf{o}_{l-1}$ is the raw output and is not quantized.

Sensitivity and calibration is performed on the same random subset of test data for each task. We use the same random seed for all results, except for ablation on sample size, where we evaluated average and standard deviation across different seeds (\ie, different samples being used in the calibration stage).

\section{Gender Personalization Results}
\label{sec:gender}

\begin{figure}[!ht]
 \centering
    \includegraphics[width=\linewidth]{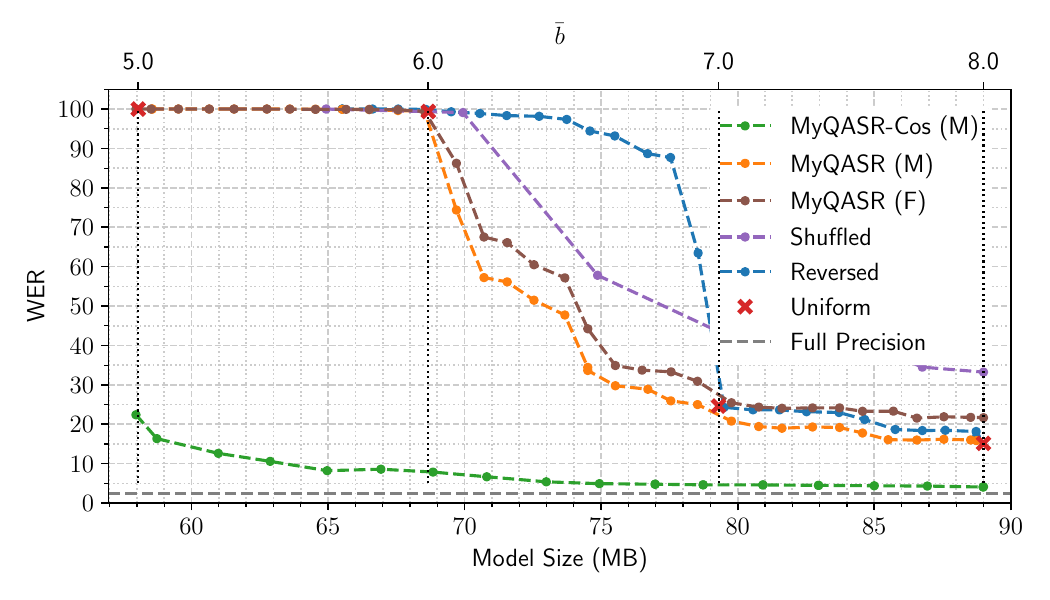}
    \caption{WER of W2V2-B model pre-trained on multi-gender data and quantized for a male user.}
    \label{fig:w2v2b_gender_male}
    
    \centering
    \includegraphics[width=\linewidth]{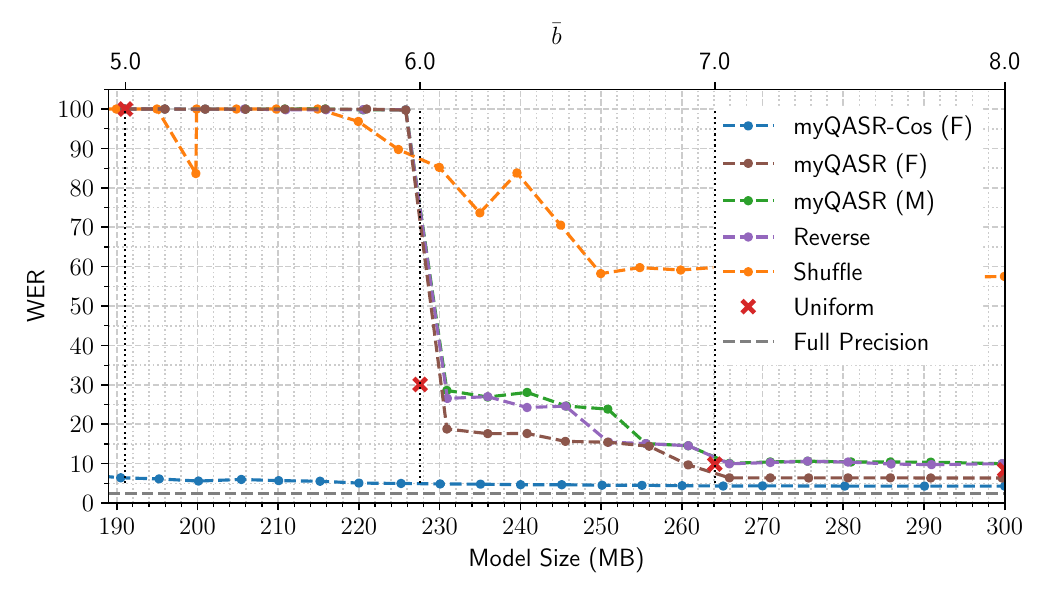}
    \caption{WER of W2V2-L model pre-trained on multi-gender data and quantized for a female user.}
    \label{fig:w2v2l_gender_female}
    
    \centering
    \includegraphics[width=\linewidth]{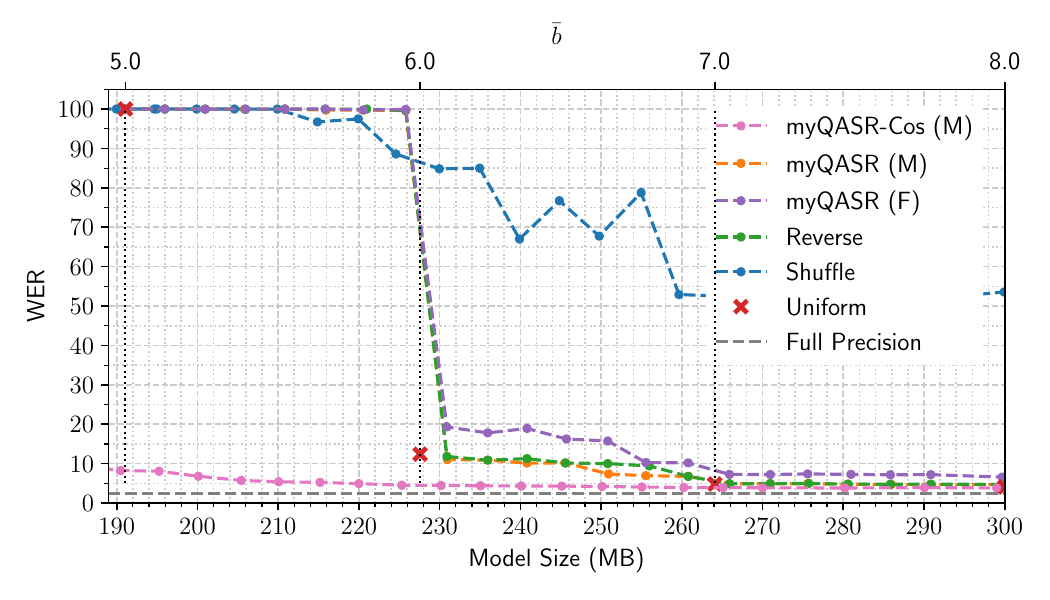}
    \caption{WER of W2V2-L model pre-trained on multi-gender data and quantized for a male user.}
    \label{fig:w2v2l_gender_male}
\end{figure}

\textbf{W2V2-B Results on Male Data.}
In the main paper we reported main results for gender-personalized model quantization on W2V2-B on female data. 
Fig.~\ref{fig:w2v2b_gender_male} shows WER of W2V2-B pre-trained on multi-gender data, quantized for a male user and tested on LS-M (symmetrical result on LS-F shown in the main paper).
The plot spans an increasing memory budget from 60MB (\ie, $\bar{b}\approx5$) to $90\mathrm{MB}$ (\ie, $\bar{b}\approx8$).
Original (\ie, FP, non-quantized) model performance indicates the lower bound for WER (gray dashed line).
Uniform quantization yields competitive results, however there is a large difference in performance when interpolating between uniform bit-depths with the reported methods. Our simplest myQASR calibrated on male data, myQASR (M), improves accuracy and can meet any desired target model size, while our more data and memory intensive method, myQASR-Cosine (M) can improve performance by a much larger margin.  
We argue on the usefulness of our sensitivity method since: 
myQASR (M) shows significant benefits compared to myQASR (F), \ie, quantizing the model according to female data; 
myQASR (M) outperforms  its shuffled (\ie, bit-depths shuffled) or reversed (\ie, bit-depths reversed) configurations.

\textbf{W2V2-L Results.}
Here, we report gender-personalized model quantization results on W2V2-L, proving the usefulness of our approach when applied to larger models. We observe that W2V-L is more robust to quantization and so we report ablation results for both $250\mathrm{MB}$ and $195\mathrm{MB}$ in the next section.
At $250\mathrm{MB}$, we observe that all calibration methods have similar performance, but as we increase the amount of compression we see how myQASR-Cosine outperforms all other calibration and sensitivity methods.

We report in Fig.~\ref{fig:w2v2l_gender_male} and Fig.~\ref{fig:w2v2l_gender_female} the WER plots for a quantized and pre-trained W2V2-L model tested on male and female data, respectively. MyQASR-Cosine performs best overall, while the more efficient myQASR method performs competitively with myQASR-Cosine when lower compression is applied. This demonstrates that it may be more appropriate in a personalized compression settings where there are computational constraints considering that myQASR-Cosine takes about $\times10$ to calibrate the model compared to myQASR.

\section{Additional Ablation Results on W2V2-L}
\label{sec:w2v2l}

In this section we report some ablation results on W2V2-L symmetric to those reported in the main paper on W2V2-B.

We analyse the components of our system one after the other one, fixing all others. W2V2-L is compressed to ${B=195\mathrm{MB}}$ where the average bit-depth is ${\bar{b}=4.5 \mathrm{bits}}$ and $B=250\mathrm{MB}$ where ${\bar{b}=5.9 \mathrm{bits}}$. We calibrate and evaluate on the same gender split. 

\noindent\textbf{Sensitivity} 
evaluation is reported in Tab.~\ref{supp:tab:ablation_sensitivity_195} and Tab.~\ref{supp:tab:ablation_sensitivity_250}.
We confirm the previous findings, outlining that median generally achieves best results and is more robust to further compression below  6 bits.

\begin{table}[t]
  \centering
  \footnotesize
  \setlength{\tabcolsep}{1.5pt}
  \caption{Ablation on the sensitivity scheme for W2V-L at 195 MB. We take only the measures used in compared methods for a fair comparison.}
  \vspace{-0.3cm}
  \resizebox{\linewidth}{!}{%
    \begin{tabular}{lcccc:lcccc}
    \toprule
    \multicolumn{1}{c}{\multirow{2}[0]{*}{\rotatebox{1.35}{\textbf{Reduction}}}} & \multicolumn{2}{c}{\textbf{Male}} & \multicolumn{2}{c}{\textbf{Female}} & \multicolumn{1}{:c}{\multirow{2}[0]{*}{\rotatebox{1.35}{\textbf{Distance}}}} & \multicolumn{2}{c}{\textbf{Male}} & \multicolumn{2}{c}{\textbf{Female}} \\
    \cmidrule(lr){2-3}\cmidrule(lr){4-5}\cmidrule(lr){7-8}\cmidrule(lr){9-10}
          & \textbf{WER} & \textbf{CER} & \textbf{WER} & \textbf{CER} &       & \textbf{WER} & \textbf{CER} & \textbf{WER} & \textbf{CER} \\
          \midrule
    Avg  & 8.4  & 2.5  &  6.6 & 2.0  & L1 \cite{yuan2022ptq4vit} & \textbf{8.0}  & \textbf{2.3}  & 8.4  & {2.6} \\
    Median (ours) & \textbf{8.0} & \textbf{2.3} & \textbf{6.1} & 1.9 & L2 \cite{yuan2022ptq4vit} & 8.2  & \textbf{2.3}  & 8.6  & 2.7 \\
    Max  & 10.5  & 3.2  & 6.7  & 2.1  & SN \cite{liu2021post} & 8.2  & \textbf{2.3}  & 8.5  & {2.6} \\
    Max Abs  & 9.5  & 3.0  & 6.3  & 1.9  & Frob \cite{9772057} & 8.2 & \textbf{2.3}  & 8.5  & 2.6 \\
    Std  & 11.1  & 3.4 & 7.0  & 2.1  & KL \cite{yuan2022ptq4vit} & 9.5  & 3.0  & 6.2  & \textbf{1.8} \\
    \bottomrule
    \end{tabular}%
    }
  \label{supp:tab:ablation_sensitivity_195}%

  \centering
  \footnotesize
  \setlength{\tabcolsep}{1.5pt}
  \caption{Ablation on the sensitivity scheme for W2V-L at 250 MB. We take only the measures used in compared methods for a fair comparison.}
  \vspace{-0.3cm}
  \resizebox{\linewidth}{!}{%
    \begin{tabular}{lcccc:lcccc}
    \toprule
    \multicolumn{1}{c}{\multirow{2}[0]{*}{\rotatebox{1.35}{\textbf{Reduction}}}} & \multicolumn{2}{c}{\textbf{Male}} & \multicolumn{2}{c}{\textbf{Female}} & \multicolumn{1}{:c}{\multirow{2}[0]{*}{\rotatebox{1.35}{\textbf{Distance}}}} & \multicolumn{2}{c}{\textbf{Male}} & \multicolumn{2}{c}{\textbf{Female}} \\
    \cmidrule(lr){2-3}\cmidrule(lr){4-5}\cmidrule(lr){7-8}\cmidrule(lr){9-10}
          & \textbf{WER} & \textbf{CER} & \textbf{WER} & \textbf{CER} &       & \textbf{WER} & \textbf{CER} & \textbf{WER} & \textbf{CER} \\
          \midrule
    Avg  & 3.1  & \textbf{0.8}  &  3.3 & \textbf{0.9}  & L1 \cite{yuan2022ptq4vit} & 4.4    & 1.3 & 4.5& 1.4 \\
    Median (ours) & \textbf{3.0} & \textbf{0.8} & \textbf{3.2} & \textbf{0.9} & L2 \cite{yuan2022ptq4vit} & \textbf{4.3}  & \textbf{1.2}  & 4.6  & 1.3 \\
    Max  & 3.1  & 0.9  & \textbf{3.2}  &  \textbf{0.9} & SN \cite{liu2021post} & 4.6  & 1.3  & 4.6  & 1.3 \\
    Max Abs  & 3.1  & 0.9  &  \textbf{3.2} & \textbf{0.9}  & Frob \cite{9772057} & 4.6 & 1.3  & 4.6  & 1.3 \\
    Std  &  3.1 & 0.9 & \textbf{3.2}  & \textbf{0.9}  & KL \cite{yuan2022ptq4vit} & 4.6  & 1.3  & 4.7  & 1.4 \\
    \bottomrule
    \end{tabular}%
    }
  \label{supp:tab:ablation_sensitivity_250}%
\end{table}

\noindent\textbf{Calibration} is evaluated in Tab.~\ref{supp:tab:ablation_calibration_195} and Tab.~\ref{supp:tab:ablation_calibration_250}, where methods with similar computation time are compared. 

Cosine and Hessian weighted myQASR perform the best while having a comparable computational search time than other approaches. 
L1 and L2 achieve competitive results at lower compression rates, however its WER is outperformed by myQASR-Hess/Cosine with a similar or lower calibration time.

\begin{table}[t]
\setlength{\tabcolsep}{4.5pt}
\footnotesize
  \centering
  \caption{Ablation on the calibration scheme for W2V-L at 195 MB. Time (seconds) measures the calibration procedure only.}
    \begin{tabular}{lcccccc}
    \toprule
    \multirow{2}[0]{*}{} & \multicolumn{3}{c}{\textbf{Male}} & \multicolumn{3}{c}{\textbf{Female}} \\\cmidrule(lr){2-4}\cmidrule(lr){5-7}
          & \multicolumn{1}{c}{\textbf{WER}} & \multicolumn{1}{c}{\textbf{CER}} & \multicolumn{1}{c}{\textbf{Time}} & \multicolumn{1}{c}{\textbf{WER}} & \multicolumn{1}{c}{\textbf{CER}} & \multicolumn{1}{c}{\textbf{Time}} \\
          \midrule
    L1 \cite{yuan2022ptq4vit} & 11.8  & 3.4  & 385   & 13.3  & 4.3  & 380 \\
    L2 \cite{yuan2022ptq4vit} & 11.15 & 3.89  & 379   & 40.7 & 18.5  & 387 \\
    LinW L2 \cite{yuan2022ptq4vit} & 36.5 & 16.7 & 378   & 77.8 & 48.2 & 388 \\
    SqW L2 \cite{yuan2022ptq4vit} & 42.5 & 20.5 & 359   & 84.5 & 60.0 & 362 \\
    myQASR-Hess & 8.8 & 2.6 & 389   & 28.9 & 12.1 & 389 \\
    myQASR-Cosine & \textbf{8.0} & \textbf{2.3} & 359   & \textbf{6.1} & \textbf{1.9} & 362  \\
    \bottomrule
    \end{tabular}%
  \label{supp:tab:ablation_calibration_195}%

\setlength{\tabcolsep}{4.5pt}
\footnotesize
  \centering
  \caption{Ablation on the calibration scheme for W2V-L at 250 MB. Time (seconds) measures the calibration procedure only.}
    \begin{tabular}{lcccccc}
    \toprule
    \multirow{2}[0]{*}{} & \multicolumn{3}{c}{\textbf{Male}} & \multicolumn{3}{c}{\textbf{Female}} \\\cmidrule(lr){2-4}\cmidrule(lr){5-7}
          & \multicolumn{1}{c}{\textbf{WER}} & \multicolumn{1}{c}{\textbf{CER}} & \multicolumn{1}{c}{\textbf{Time}} & \multicolumn{1}{c}{\textbf{WER}} & \multicolumn{1}{c}{\textbf{CER}} & \multicolumn{1}{c}{\textbf{Time}} \\
          \midrule
    L1 \cite{yuan2022ptq4vit} & 3.4  & 0.9  & 102   & 3.6  & 1.0  & 101 \\
    L2 \cite{yuan2022ptq4vit} & 3.0 & 0.9  & 99   & 3.2 & 1.0  & 102 \\
    LinW L2 \cite{yuan2022ptq4vit} & 3.0 & 0.9 & 103   & 3.2 & 1.0 & 100 \\
    SqW L2 \cite{yuan2022ptq4vit} & 3.3 & 0.9 & 103   & 3.3 & 1.0 & 100 \\
    myQASR-Hess & 3.0 & 0.8 & 104  & 3.3 & 0.9 & 105 \\
    myQASR-Cosine & \textbf{3.0} & \textbf{0.8} & 112   & \textbf{3.2} & \textbf{0.9} & 113  \\
    \bottomrule
    \end{tabular}%
  \label{supp:tab:ablation_calibration_250}%
\end{table}%

\section{Qualitative Model Transcriptions}
\label{sec:qualitative}

We report in Tab.~\ref{tab:text-examples} some sample qualitative results in terms of output model transcriptions to appreciate the usefulness of the proposed approach. In the table, we compress W2V2-B model trained on multi-gender data and then compressed using our method for both male (M) and female users (F). We then evaluate on female users to show how our personalized compression method (myQASR) improves the ASR results. 

\begin{table*}[t] 
\setlength{\tabcolsep}{6pt}
\scriptsize
  \centering
  \caption{Examples of transcription outputs of myQASR on W2V2-B quantized to 75 MB.}
  \vspace{0cm}
    \begin{tabular}{p{0.3\linewidth} p{0.3\linewidth} p{0.3\linewidth}}
  \toprule
\textbf{Calibration:M Test:F} & \textbf{Calibration:F Test:F} & \textbf{Ground Truth}\\ \midrule
ALSO A POPULAR CONTRIVANCE WHEREBY LOVE   MAKEN MAY BES A SPENDID MBUT NOT STOPED T TPCSI & ALSO A POPULAR CONTRIVANCE WHEREBY LOVE MAKING MAY BE SUSPENDID BUT NOT   STOPPED DURING THE TECNIXESON & ALSO A POPULAR CONTRIVANCE WHEREBY LOVE   MAKING MAY BE SUSPENDED BUT NOT STOPPED DURING THE PICNIC SEASON \\ \midrule
HARANGUE THE TIRESOME PHLODICT TOF THES   TIRELESS TON & HARANGUE THE TIRESOME TRODUCT OF A TIRELESS TONGUE & HARANGUE THE TIRESOME PRODUCT OF A   TIRELESS TONGUE \\\midrule
ANGOR HA THANKFUL TO HEAR & ANGOR HAIN PAINFUL TO HEAR & ANGOR PAIN PAINFUL TO HEAR \\\midrule
HAY FEVER A HARD TROUBLE FILES FI FALLING   IN LOVE WITH THE GRASP WITHER & HAY FEVER A HEART TROUBLE CAUSE FY FALLING IN LOVE WITH A GRASS WIDOW & HAY FEVER A HEART TROUBLE CAUSED BY   FALLING IN LOVE WITH A GRASS WIDOW \\\midrule
HEAVEN AND GOOD PLACE TO BE RAISED TO & HEAVEN A GOOD PLACE TO BE RAISED TO & HEAVEN A GOOD PLACE TO BE RAISED TO \\\midrule
HECH OFFENCE & KEDGE AFFENCE & HEDGE A FENCE \\\midrule
FEREDITY THE CAUSE OF ALL OUR FAULTS & HEREDITY THE CAUSE OF ALL OUR FAULTS & HEREDITY THE CAUSE OF ALL OUR FAULTS \\\midrule
CORE SSE A DEGREE OF WISDOM THAT KEEPS   ONE FROM BETTING ON THE RACES & FORSE SENSE A DEGREE OF WISDOM THAT KEEPS ONE FROM BETTING ON THE RACES & HORSE SENSE A DEGREE OF WISDOM THAT KEEPS   ONE FROM BETTING ON THE RACES \\\midrule
HOVE MAS EXCUSE FOR WETTING THE WALK & POSE MAN'S EXCUSE FOR WETTING THE WALK & HOSE MAN'S EXCUSE FOR WETTING THE WALK \\\midrule
HOTEL A PLACE WHERE A GUEST OFTEN GIVES   UP GOOD DOLLARS FOR  POR PORTERS & HOTEL A PLACE WHERE A GUEST OFTEN GIVES UP GOOD DOLLARS FOR POR QUARTERS & HOTEL A PLACE WHERE A GUEST OFTEN GIVES   UP GOOD DOLLARS FOR POOR QUARTERS \\\midrule
HOUSELAD A DOMESTIC OF HEEPLEER  THAT MAKES IC EASY FOR THE GOVERENT TO   UNLOST ALL THE SOULTS  NEEDS & HOUSE CLEANING A DOMESTIC OF HEABL THAT MAKES IT EASY FOR THE GOVERNMENT   TO ENLIST ALL THE SOLDIERS IT NEEDS & HOUSECLEANING A DOMESTIC UPHEAVAL THAT   MAKES IT EASY FOR THE GOVERNMENT TO ENLIST ALL THE SOLDIERS IT NEEDS \\\midrule
OSBIC    TTHE NEXT THING TO A WIFE & HUSBAND TTHE NEXT THING TO A WIFE & HUSBAND THE NEXT THING TO A WIFE \\\midrule
FUSSY WOMAN AND LONG TIE & HUSSY WOMAN AND BAN TY & HUSSY WOMAN AND BOND TIE \\\midrule
TIED    TO A WOMAN & TIED TO A WOMAN & TIED TO A WOMAN \\\midrule
HYPOCRATE BAFLE STEALER & GIVE A CRIT A HORSE TEALER & HYPOCRITE A HORSE DEALER \\\midrule
THOSE PRETTY WRONGS TAT LIBERTY TOOFTI   WHEN I AM SOMETIME ABSENT FROM TY HART    THY BEAUTY AND THY YEARS FULL WELL BEFICS D FORSTILL TEMPTATION   FOLLOWED WHERE THOU ART & THOSE PRETTY WRONGS THAT LIBERTY COMMITS WHEN I AM SOMETIME ABSENT FROM   THY HART THY BEAUTY AND THY YEARS FULL WELL BEFITS FORSTILL TEMPTATION   FOLLOWS WHERE THOU ART & THOSE PRETTY WRONGS THAT LIBERTY COMMITS   WHEN I AM SOMETIME ABSENT FROM THY HEART THY BEAUTY AND THY YEARS FULL WELL   BEFITS FOR STILL TEMPTATION FOLLOWS WHERE THOU ART \\\midrule
I BE & A MEAN & AY ME \\\midrule
NO MATTER GEND ALTHOUGH MY FOOT DID STAND   UPON THE FARTHEST EARTH REMOVED FROM DHE FOR NIMBLE THOUGHT  IN JUMP    BOTH FEE AND LAND AS SOON AS THING D     LAC  WER HE WOULD BE BUT AH & NO MATTER THEN ALTHOUGH MY FOOT DID STAND UPON THE FARTHEST EARTH REMOVED   FROM THEE FOR NIMBLE FOUGHT CAN JUMP OTH SEE AND LAND AS SOON AS THINK THE   PLACE WHERE HE WOULD BE BUT AH & NO MATTER THEN ALTHOUGH MY FOOT DID STAND   UPON THE FARTHEST EARTH REMOV'D FROM THEE FOR NIMBLE THOUGHT CAN JUMP BOTH   SEA AND LAND AS SOON AS THINK THE PLACE WHERE HE WOULD BE BUT AH \\\midrule
THOT KILLS ME THAT I AM NOT BUT TO LEADE   LARGE LENGTHS O MILES THOU ART GONE BUT THAT SO MUCH OF EARTH AND WATER WROUT   I MUST ATTEND TIMES LEISURE WITH MY MON RECEIVING NOT BY ELEMENT SOTSLOW BUT   HEAVY TEARS BAGGERS RFITHERS WNE & THOUGHT KILLS ME THAT I AM NOT BHOUGT    TO LEAP LARGE LENGTHS MILES WHEN THOU ART GONE BUT THAT SO MUCH OF   EARTH AND WATER WROUGT I MUST ATTENDS TIMES LEISURE WITH MY MOAN H RECEIVING   NOT BY ELEMENTS SO SLOW BUT HEAVY TEARS ADGES OF EITHER'S WOE & THOUGHT KILLS ME THAT I AM NOT THOUGHT TO   LEAP LARGE LENGTHS OF MILES WHEN THOU ART GONE BUT THAT SO MUCH OF EARTH AND   WATER WROUGHT I MUST ATTEND TIME'S LEISURE WITH MY MOAN RECEIVING NOUGHT BY   ELEMENTS SO SLOW BUT HEAVY TEARS BADGES OF EITHER'S WOE \\\midrule
MY HEART DOT PLEED THAT THOU INHIM DOFF   EMINE ACLA THAT NEVER PIERCED O FRYSTAL MINE BUT TE DEFENDEC DOFT THAT FE   DENIN AND TAS IN HIM THY FAIR FER INMINE & MY HART DOTH PLEAD THAT THOU IN HYM DOTH LIE A CLOSET NEVER PIERCED WITH   CRYSTAL IES BUT THE DEFENDANT DOTH THAT PLEAD DENIE AND SAYS IN HYM THY FAIR   APPARANCE LIES & MY HEART DOTH PLEAD THAT THOU IN HIM DOST   LIE A CLOSET NEVER PIERC'D WITH CRYSTAL EYES BUT THE DEFENDANT DOTH THAT PLEA   DENY AND SAYS IN HIM THY FAIR APPEARANCE LIES \\\midrule
YOU ARE MY ALL THE WORLD  I MUST STRIVE TO KNOW MY SHAMES AND PRASES   FROM OR TON NONE ELSE TO ME NOR I TO NON ALIVE WITH MY STEELED SENSE OR   CHANGES RIGHT OR WRONG & YOU ARE MY ALL THE WORLD AND I MUST STRIVE TO KNOW MY CHAMES AND PRAISES   FROM YOUR TONGUE NONE ELSE TO NEED NOR I TO NONE ALIVE THAT NMY STEALED SENSE   OR CHANGES RIGHT OR WRONG & YOU ARE MY ALL THE WORLD AND I MUST   STRIVE TO KNOW MY SHAMES AND PRAISES FROM YOUR TONGUE NONE ELSE TO ME NOR I   TO NONE ALIVE THAT MY STEEL'D SENSE OR CHANGES RIGHT OR WRONG \\\midrule
OH TIS THE FIRST TIS FLATTERY IN MY SE   AND MY GREAT MIND MOST KINGLY DRINKS AT O MINE IYE WELL NOWSWLOT WITH IS   GOFFIS IS REE AND TO HIS PALET DOCPOPAIR THE COCK IF IT BE POISONTIS THE   LASSER SIN THAT MINE I LOVSN'TAN DORK FOST BEGIN & OH TIS THE FIRST TS FLATTERY IN MY SEEN AND MY GREAT MIND MOST KINGLY   DRINKS IT OFF MINE EYE WELL KNOWS WHAT WITH HIS GUST IS GREEN AND TO HIS   PALLAT DO PREPARE THE CUF IF IT BE POISONED TIS THE LESSER SIN  LET MINE EYE LOVES IT  AND OFT FIRST BEGIN & O TIS THE FIRST TIS FLATTERY IN MY SEEING   AND MY GREAT MIND MOST KINGLY DRINKS IT UP MINE EYE WELL KNOWS WHAT WITH HIS   GUST IS GREEING AND TO HIS PALATE DOTH PREPARE THE CUP IF IT BE POISON'D TIS   THE LESSER SIN THAT MINE EYE LOVES IT AND DOTH FIRST BEGIN \\\midrule
BUT RECKONING TIME WISH MILLIOND   ACCIDENTS PREP INTWIXST FOWS ANDCHANGE DECREES OF TENS HAND SACRED DUTY BLUNK   TO THE SHARKSTINTENSE DIVERT STRONG MINDS TO THE FORCE OF ALTEREN IN ALAS WY   FEARING OF TIMES TERANY MIGHT I NOT THEN SAY NOW I LOVE BEST WHEN I WITH   CERTAIN OR INCERTAINTY FROWNING PRESENT DOUBTING THE BREST & BUT RECKONING TIME WHOSE MILLIONED ACCIDENTS CREEP IN TWIXT VOWS AN   CHANGE DECREES OF KINGS TANDS SACRED BEAUTY BLUNT THE SHARPEST AND TENSS   DIVERT STRONG MINDS TO THE COURSE OF ALTERING THINGS ALAST WHY FEARING OF   TIMES TYRANNY MIGHT I NOT THEN SAY TNOW I LOVE YOU BEST WHEN I WAS CERTAIN OR   IN CERTAINTY CROWNING OF PRESENT DOUBTING OF THE REST & BUT RECKONING TIME WHOSE MILLION'D   ACCIDENTS CREEP IN TWIXT VOWS AND CHANGE DECREES OF KINGS TAN SACRED BEAUTY   BLUNT THE SHARP'ST INTENTS DIVERT STRONG MINDS TO THE COURSE OF ALTERING   THINGS ALAS WHY FEARING OF TIME'S TYRANNY MIGHT I NOT THEN SAY NOW I LOVE YOU   BEST WHEN I WAS CERTAIN O'ER INCERTAINTY CROWNING THE PRESENT DOUBTING OF THE   REST \\ \bottomrule
    \end{tabular}%
  \label{tab:text-examples}%
  \vspace{0cm}
\end{table*}

\end{document}